\documentclass[10pt]{article}

\usepackage{amsmath}
\usepackage{amssymb}
\usepackage{graphicx}
\usepackage{url}
\usepackage{booktabs}
\usepackage{color}
\usepackage[table]{xcolor}

\begin{document}

\title{\textbf{THE ROLE OF TASTE AFFINITY IN AGENT-BASED MODELS FOR SOCIAL RECOMMENDATION}}
\author{Giulio Cimini\footnote{giulio.cimini@unifr.ch}, An Zeng, Mat\'u\u{s} Medo\\ \emph{Physics Department, University of Fribourg, CH-1700 Fribourg, Switzerland} \vspace{0.4cm}\\
Duanbing Chen\\ \emph{Web Sciences Center, UESTC, Chengdu 611731, People's Republic of China}}
\date{}

\maketitle


\begin{abstract}
In the Internet era, online social media emerged as the main tool for sharing opinions and information among individuals. 
In this work we study an adaptive model of a social network where directed links connect users with similar tastes, 
and over which information propagates through social recommendation. 
Agent-based simulations of two different artificial settings for modeling user tastes are compared with patterns seen in real data, 
suggesting that users differing in their scope of interests is a more realistic assumption than users differing only in their particular interests. 
We further introduce an extensive set of similarity metrics based on users' past assessments, 
and evaluate their use in the given social recommendation model with both artificial simulations and real data. 
Superior recommendation performance is observed for similarity metrics that 
give preference to users with small scope---who thus act as selective filters in social recommendation.
\end{abstract}

\textbf{Keywords}: Taste similarity, agent-based modeling, adaptive complex networks, information diffusion and filtering, social recommendation.

\section{Introduction}\label{sec.intro}

Online social networks (Facebook, Twitter, Digg, and others) are nowadays popular media which allow opinions and information to spread rapidly to friends, friends of friends, 
and further, eventually reaching a huge audience by cascading chains of influences \cite{BrownJCR1987,KempeSIGKDD2003}. 
Furthermore, they are able to reflect moods and current beliefs of people, giving early signs of what is just going to happen \cite{BollenJCS2011,AsurIEEE2010}. 
The study of the structure and evolution of social networks has hence attracted much more attention than in the past, becoming a main focus of complexity science. 
To this end, agent-based models (ABM) represent an ideal tool for simulating the social behavior of autonomous individuals and the interactions among them \cite{AxelrodSSS1997}, 
and have been successfully used to model decentralized scenarios where individual interactions lead to the emergence of collective patterns \cite{HamillJASS2009,SingerPRE2009}. 
When modeling social networks however one should be aware that they differ from other kind of complex systems because of two features: 
\emph{taste affinity} and \emph{social influence} between users. 
Taste affinity (also named homophily or selection in the literature) is the tendency of individuals to form relationships with others who posses 
similar characteristics \cite{McPersonARS2001}. Social influence is the process which leads people to adopt behaviors exhibited by those they interact with \cite{FriedkinCUP1998}. 
Taste affinity and social influence not surprisingly lead to correlation between actions of adjacent nodes in the network, 
and hence are nowadays widely recognized to play a fundamental role in the formation and evolution of social systems \cite{KumarSIGKDD2008,CrandallSIGKDD2008}.
\newline\newline

In this work we focus on taste affinity between users. While past studies of homophily in online social network have been based on the analysis of emotionally annotated data 
\cite{MitrovicEPJB2010,ChmielPLOS2011,GarasSCR2012}, here we build our study on the adaptive model of social recommendation introduced in \cite{MedoEPL2009,WeiPhysA2011}. 
The model consists of a network of users, each connected to other users who represent her information sources---which we name the user's \emph{leaders}. 
Users in the community interact with some kind of content---we use news here, but they can be also movies, books, bookmarks, and so on. 
The propagation of a news over the network starts from the user who uploads it to the system and then automatically forwards it to her \emph{followers}---the users who have chosen her as leader. 
If one of these followers later likes the news, it will be similarly passed further to this user's followers, and so on. 
This social spreading/filtering mechanism is known as \emph{social recommendation}, 
and it is based on the evidence that users prefer recommendations made by friends rather than those by a centralized system \cite{SinhaDELOSNSF2001,BorgattiMS2003}. 
Simultaneously with the propagation of news, connections of the leader-follower network can change when, for any user, a leader better than her current ones is found. 
The quality of the candidate leader is estimated by comparing the users' past assessments to obtain their \emph{taste similarity}. Due to this mechanism, 
the evolution of the social network and the dynamical processes taking place on the network become coupled, as it happens in real social communities \cite{AgarwalISR2008,BampoISR2008}. 
This gives rise to a rich variety of features, for instance the emergence of a ``good get richer'' mechanism which leads to scale-free leadership structures---again 
observed in real social networks \cite{CiminiEPJB2011,ZhouPLoS2011}. 
\newline\newline
To study the evolution and dynamics of the system, we first turn to computer simulations and introduce a general agent-based framework for describing users' tastes. 
We build on the assumption that users' tastes and news' attributes can be represented by vectors, 
and that the assessment of a user on a news depends on the overlap between the respective vectors. 
A fundamental issue is hence how to construct these vectors. Usually one takes inspiration from real data, 
where dimensionality reduction techniques (Singular Value Decomposition, Probabilistic Latent Semantic Analysis and Latent Dirichlet Allocation) 
can be employed to extract analogous features from the given data \cite{KorenIEEE2009}. 
However the result of this approach strongly depends both on the number of used hidden variables and, more importantly, on the particular dataset analyzed. 
Here we take the opposite approach: instead of seeking for a perfect model of users' interests, we study the system under two different plausible models of users' tastes. 
We then apply our social recommendation model to real data and determine which taste model better describes reality. 
\newline\newline
The way of modeling users' tastes determines users' actions and, at last, the behavior of the whole system. 
However taste vectors are hidden variables both in our model and in reality. 
We hence tackle the opposite issue as well: given the users' actions, how to quantify users' taste similarity?
In the literature, there are a number of definitions for taste similarity based on users' assessment: 
the probability of agreement, the cosine similarity, the Jaccard coefficient, the Pearson correlation coefficient, to give a few examples, 
and each definition is apt for a particular setting. For instance, Jaccard's works with implicit ratings, whereas Pearson's exploits explicit assessments. 
Assuming that users may like, dislike or not rate an item, we make a comparative study of a number of similarity metrics. 
We use both simulations and real data to determine which of them are better for building an efficient leader-follower network for social recommendation.  
We show that using a well-chosen definition of the similarity can significantly enhance the quality of the produced recommendations, and ultimately improve users' satisfaction. 
\newline\newline
To summarize, our contribution is twofold: 
1) we study two different ABMs for social systems built upon plausible assumptions about users' tastes, determining which one is closer to reality, and 
2) we identify a set of similarity metrics which are suitable to measure taste affinity, and which can be used to obtain quality recommendations 
both in simulations (under different ABM assumptions) and in real case studies. 
The rest of the work is organized as follows. In Section 2 we describe the model for social recommendation along with different ways of modeling users' tastes 
and measuring taste similarity. Section 3 contains results of agent-based simulations, whereas Section 4 presents results obtained on real datasets. 
Conclusions are given in Section 5.

\section{The Model}\label{sec.model}

We now describe the social recommendation model originally introduced in \cite{MedoEPL2009}.

\emph{Notation.} We denote as $U$ the total number of users, $L$ the number of leaders assigned to each user 
and $s_{i|j}$ the estimated similarity of reading tastes of user $i$ with respect to those of user $j$---obtained from $i$'s and $j$'s assessments of news 
(see Sec. \ref{sec.similarity}). Evaluation of news $\alpha$ by user $i$ ($e_{i\alpha}$) can be either $+1$ (liked), $-1$ (disliked) or $0$ (not rated yet). 
The model assumes no other information than ratings of news by users.

\emph{Network structure.} 
The set of $U$ users is represented as a monopartite social network, in which each user $i$ is connected by directed links to $L$ other users 
who represent $i$'s leaders (information sources). Conversely, users who have $i$ as a leader are referred to as followers of $i$. 
Thus, a directed edge points from $j$ to $i$ ($j\rightarrow i$) if user $j$ is a leader of user $i$ or, equivalently, $i$ is a follower of $j$.

\emph{News spreading.}
The directed leader-follower network serves as the basis for news propagation. 
When user $k$ introduces news $\alpha$ to the system, the news is automatically passed to all followers of $k$. 
For follower $j$, the \emph{social recommendation score} of news $\alpha$ is set proportional to $j$'s similarity with the news's submitter: $R_{j\alpha}\propto s_{j|k}$ 
(the higher the $s_{j|k}$, the stronger the recommendation). If $\alpha$ is later approved by user $j$, it is similarly passed further to all users $i$ who are followers of $j$ 
with recommendation score proportional to their similarity with $j$: $R_{i\alpha}\propto s_{i|j}$. 
If $\alpha$ is approved by one of users $i$, it is passed further to $i$'s followers, and so forth. 
Note that if a user receives the same news from multiple leaders, the news's recommendation score becomes proportional to the sum of the similarity scores between this user 
and the leaders who forwarded it, reflecting that a news approved by several leaders is more likely to be approved by this user too.
Finally, to allow fresh news to be accessed fast, recommendation scores are damped with time with a function $(1-\tau^{-1})^{t-t_\alpha}$, 
where $t$ represents the current time, $t_\alpha$ is the time news $\alpha$ was introduced in the system, and $\tau$ is the time scale of the decay. 
Summing up, for a generic user $i$ at time $t$ the recommendation score for news $\alpha$ is: 
\begin{equation}\label{eq.RS}
R_{i\alpha}(t)=(1-\tau^{-1})^{t-t_\alpha}\,\sum_{j\in L_i}s_{i|j}\:\delta_{e_{j\alpha},1}
\end{equation}
Here $L_i$ is the set of leaders of user $i$ and the term $\delta_{e_{j\alpha},1}$ is one when user $j$ liked news $\alpha$ ($e_{j\alpha}=1$) and zero otherwise 
($\delta$ is the Kronecker symbol). Recommendation scores are used to sort the \emph{unread} news each user receives (high scores at the top).

\emph{Leaders updating.} 
Simultaneously with the propagation of news, connections of the leader-follower network are rewired when, for any user, a leader better than her current ones is found. 
In practice this means that if $j$ is the leader of $i$ with the lowest similarity value, 
$j$ is replaced with a new user $k$ if $s_{i|k}>s_{i|j}$ (link $j\rightarrow i$ is cut and $k\rightarrow i$ is created). 
As selection strategy for picking new candidate leaders, we employ a so-called \emph{Hybrid rewiring} \cite{CiminiPRE2012}, which is composed 90\% of $LL+F$ \emph{local rewiring} 
($k$ is chosen from the leaders' leaders and the followers of user $i$ such that $s_{i|k}$ is maximized) and 10\% of \emph{random rewiring} ($k$ is simply a user picked at random from all users). 
This rewiring mechanism mimics the natural evolution of social communities where users search for friends among friends of friends 
(local rewiring) but also casual encounters occur and may lead to long-term relationships (random rewiring). 
By construction, the Hybrid rewiring naturally enhances both the link reciprocity \cite{Wasserman-CUP-1994} and the clustering coefficient \cite{Fagiolo-PRE-2007} of the social network \cite{CiminiPRE2012}.
In addition, it was shown to represent an optimal compromise between complexity and efficiency \cite{CiminiPRE2012}. 
Of course, there are more sophisticated models of network evolution, which are based on statistical methods to estimate the model's parameters from empirical data 
\cite{LeskovecSIGKDD2008,SnijdersSN2010}. However, since it is not our goal to model a particular network, 
we stick on simple and general evolution rules---which nevertheless were shown to produce network topologies 
with high values of link reciprocity and clustering coefficient, which thus closely resemble the topologies of real social communities \cite{CiminiPRE2012}. 

\subsection{User Similarity}\label{sec.similarity}

An essential ingredient of the model is the estimated similarity of users' reading tastes, which regulates the news' flow over the system 
by determining the leaders' selection from users (i.e., the link structure of the network) and recommendation scores of news. 
Since we can rely only on users' ratings, the similarity of a pair of users has to be estimated from their assessments on news---which in our case 
can be either like, dislike, or not rated. 
\newline\newline
The first similarity measure that we consider, first introduced in \cite{MedoEPL2009}, is the overall probability of agreement between users $i$ and $j$. 
If $l_i=\{\alpha|e_{i\alpha}=1\}$ and $d_i=\{\alpha|e_{j\alpha}=-1\}$ are respectively the set of news liked and disliked by user $i$, then:
\begin{equation}\label{eq.s0}
s_{i|j}^{(0)}:=\frac{|l_i \bigcap l_j|+|d_i \bigcap d_j|}{|l_i \bigcap l_j|+|d_i\bigcap d_j|+|d_i\bigcap l_j|+|l_i\bigcap d_j|}
\end{equation}
where the denominator is the total number of news read by both $i$ and $j$. 
We then consider the Jaccard coefficient, which however only works with binary ratings. 
By considering only like instances (i.e., disregarding dislikes) we get: 
\begin{equation}\label{eq.j0}
J_{i|j}^{(0)}:=\frac{|l_i \bigcap l_j|}{|l_i \bigcup l_j|}
\end{equation}
which represent the ratio of the number of news liked by both $i$ and $j$ over the number of news liked by either of them. 
The opposite procedure (disregarding likes) gives:
\begin{equation}\label{eq.jt0}
\tilde{J}_{i|j}^{(0)}:=\frac{|d_i \bigcap d_j|}{|d_i \bigcup d_j|}
\end{equation}
We have tested this similarity and obtained poor results (not reported in this paper), which is not surprising. 
The social recommendation mechanism is in fact based on users' filtering: leaders forward to their followers only the news they like. 
This means that a news liked by leader $j$ is always passed to follower $i$, who will possibly evaluate it. 
However if $j$ does not like a news, $i$ does not receive it directly\footnote{In principle, $i$ could still receive the news from another leader; 
however, such leader has to receive the news first, and then like and forward it. This longer path is hindered both by the time-damping of recommendation scores 
and by the eventuality that leaders of the same user have similar tastes among themselves.} and it is less likely that $i$ will ever evaluate it. For this reason, 
the sets $|l_i \bigcap l_j|$ and $|d_i \bigcap l_j|$ are much larger than $|d_i \bigcap d_j|$ and $|l_i \bigcap d_j|$, 
and a similarity score based on either of the first sets is much more reliable in measuring taste affinity than a score based on either of the second sets. 
Hence in what follows we will not consider measures based only on $|d_i \bigcap d_j|$ or $|l_i \bigcap d_j|$. 
\newline\newline
Coming back to $J^{(0)}$, we do not expect it to work better than $s^{(0)}$ because it exploits less information. 
We instead use $J^{(0)}$ as a baseline for introducing other similarity metrics:
\begin{eqnarray}
J_{i|j}^{(1)}&:=&\frac{|l_i \bigcap l_j|}{|l_i \bigcup l_j|}+\frac{|d_i \bigcap d_j|}{|d_i \bigcup d_j|}\label{eq.j1}\\
J_{i|j}^{(2)}&:=&\frac{|l_i \bigcap l_j|}{|l_i \bigcup l_j|}-\frac{|d_i \bigcap l_j|}{|d_i \bigcup l_j|}\label{eq.j2}\\
J_{i|j}^{(3)}&:=&\frac{|l_i \bigcap l_j|}{|l_i \bigcup l_j|}-\frac{|l_i \bigcap d_j|}{|l_i \bigcup d_j|}\label{eq.j3}\\
\end{eqnarray}
Note that $s^{(0)}$, $J^{(0)}$ and $J^{(1)}$ are symmetric measures: $s_{i|j}^{(0)}\equiv s_{j|i}^{(0)}$, $J_{i|j}^{(0)}\equiv J_{j|i}^{(0)}$ and $J_{i|j}^{(1)}\equiv J_{j|i}^{(1)}$. 
Instead $J^{(2)}$ and $J^{(3)}$ are not symmetric---what happens is that $J_{j|i}^{(2)}\equiv J_{i|j}^{(3)}$ and vice versa. 
Since our network is directed (allowing for situations where $j$ is a good leader for $i$ but not the contrary), 
considering asymmetric similarity scores is more than appropriate. We pursue this direction further by introducing asymmetric equivalents of $s^{(0)}$ and of $J^{(0)}$:
\begin{eqnarray}
s_{i|j}^{(0L)}&:=&\frac{|l_i \bigcap l_j|+|d_i \bigcap d_j|}{|l_j|+|d_j|}\label{eq.s0l}\\
s_{i|j}^{(0F)}&:=&\frac{|l_i \bigcap l_j|+|d_i \bigcap d_j|}{|l_i|+|d_i|}\label{eq.s0f}\\
J_{i|j}^{(0L)}&:=&\frac{|l_i \bigcap l_j|}{|l_j|}\label{eq.j0l}\\
J_{i|j}^{(0F)}&:=&\frac{|l_i \bigcap l_j|}{|l_i|}\label{eq.j0f}
\end{eqnarray}
Here $s_{i|j}^{(0L)}$ ($s_{i|j}^{(0F)}$) is the probability of follower $i$ and leader $j$ agreeing on the news assessed by leader $j$ (follower $i$), 
whereas $J_{i|j}^{(0L)}$ ($J_{i|j}^{(0F)}$) is the probability of follower $i$ (leader $j$) liking a news liked by leader $j$ (follower $i$). 
Asymmetry here translates into $s_{j|i}^{(0L)}\equiv s_{i|j}^{(0F)}$ and $J_{j|i}^{(0L)}\equiv J_{i|j}^{(0F)}$.
\newline\newline
We finally consider asymmetric measures which, as for $J^{(1)}$, $J^{(2)}$ and $J^{(3)}$, combine the sets $l_i$, $l_j$, $d_i$, $d_j$ as separate algebraic contributions. 
There are many possibilities to construct such similarities but as we said before, news liked by leader $j$ are the ones which are recommended to follower $i$:
it makes more sense to use the probability of $i$ liking a news liked by $j$. For this reason, we build on $J_{i|j}^{(0L)}$ and introduce:
\begin{eqnarray}
K_{i|j}^{(1L)}&:=&\frac{|l_i \bigcap l_j|}{|l_j|}+\frac{|d_i \bigcap d_j|}{|d_j|}\label{eq.k1l}\\
K_{i|j}^{(1F)}&:=&\frac{|l_i \bigcap l_j|}{|l_j|}+\frac{|d_i \bigcap d_j|}{|d_i|}\label{eq.k1f}\\
K_{i|j}^{(2L)}&:=&\frac{|l_i \bigcap l_j|}{|l_j|}-\frac{|d_i \bigcap l_j|}{|l_j|}\label{eq.k2l}\\
K_{i|j}^{(2F)}&:=&\frac{|l_i \bigcap l_j|}{|l_j|}-\frac{|d_i \bigcap l_j|}{|d_i|}\label{eq.k2f}\\
K_{i|j}^{(3L)}&:=&\frac{|l_i \bigcap l_j|}{|l_j|}-\frac{|l_i \bigcap d_j|}{|d_j|}\label{eq.k3l}\\
K_{i|j}^{(3F)}&:=&\frac{|l_i \bigcap l_j|}{|l_j|}-\frac{|l_i \bigcap d_j|}{|l_i|}\label{eq.k3f}
\end{eqnarray}
The meaning of each definition is straightforward: $K_{i|j}^{(1L)}$ is equal to $J_{i|j}^{(0L)}$ plus the probability of follower $i$ disliking a news disliked by leader $j$; 
$K_{i|j}^{(2F)}$ is equal to $J_{i|j}^{(0L)}$ minus the probability of leader $j$ liking a news disliked by follower $i$; and so forth. 
We remark that this list of similarities is not intended to be exhaustive.
\footnote{There are many other ways to assess users' similarities from the user-object bipartite graph and the user-user social graph: the generalized model of relational similarity \cite{KovacsSN2010}, 
blockmodeling \cite{BorgattiSM1992}, the CONCOR algorithm \cite{BreigerJMP1975}, automorphic equivalence \cite{GreenacreTACA1984}, Correspondence Analysis \cite{WhiteAJS1976}, and many others. 
However, these methods require as input the whole dataset of users and objects, also to evaluate the similarity of a single user pair, and are much demanding in terms of computational cost. 
For this reason, they are not applicable in practice to online social networks with millions of users and continuously generated content---and thus have not been considered in our analysis.}
Instead, we considered various definitions containing all the combinations of the sets $l_i$, $l_j$, $d_i$, $d_j$ 
which are meaningful in our opinion and have the potential to achieve a reasonably good performance. 
\newline\newline
Table \ref{tab.summS} gives a summary of the similarity metrics that we introduced above.
Note that all definitions of the similarity are prone to statistical fluctuations when the set of news considered for the assessment is small, 
i.e. when the denominator of an algebraic term is small. Since in a sampling of $n$ trials the typical relative fluctuation is of the order of $1/n$, 
we multiply each separate term of all introduced similarities by a factor $(1-n^{-1/2})$, where $n$ is the denominator of the considered term. 
Such factor disadvantages user pairs with a small $n$. Finally, when we do not have enough evaluation for a user pair (meaning that their similarity is undefined) 
we set it to a small positive value $s_0$, reflecting the fact that even when there are no users' evaluations, there is some base similarity of their interests.

\begin{table}
\centering
 \caption{\scriptsize{List of similarity metrics.}\label{tab.summS}}
 \begin{tabular}{cl}
symbol	& formula	\\
\toprule
$s_{i|j}^{(0)}$&$\dfrac{|l_i \bigcap l_j|+|d_i \bigcap d_j|}{|l_i \bigcap l_j|+|d_i\bigcap d_j|+|d_i\bigcap l_j|+|l_i\bigcap d_j|}$\\
$J_{i|j}^{(0)}$&$\dfrac{|l_i \bigcap l_j|}{|l_i \bigcup l_j|}$\\
\midrule
$J_{i|j}^{(1)}$&$\dfrac{|l_i \bigcap l_j|}{|l_i \bigcup l_j|}+\dfrac{|d_i \bigcap d_j|}{|d_i \bigcup d_j|}$\\
$J_{i|j}^{(2)}$&$\dfrac{|l_i \bigcap l_j|}{|l_i \bigcup l_j|}-\dfrac{|d_i \bigcap l_j|}{|d_i \bigcup l_j|}$\\
$J_{i|j}^{(3)}$&$\dfrac{|l_i \bigcap l_j|}{|l_i \bigcup l_j|}-\dfrac{|l_i \bigcap d_j|}{|l_i \bigcup d_j|}$\\
\midrule
$s_{i|j}^{(0L)}$&$\dfrac{|l_i \bigcap l_j|+|d_i \bigcap d_j|}{|l_j|+|d_j|}$\\
$s_{i|j}^{(0F)}$&$\dfrac{|l_i \bigcap l_j|+|d_i \bigcap d_j|}{|l_i|+|d_i|}$\\
$J_{i|j}^{(0L)}$&$\dfrac{|l_i \bigcap l_j|}{|l_j|}$\\
$J_{i|j}^{(0F)}$&$\dfrac{|l_i \bigcap l_j|}{|l_i|}$\\
\midrule
$K_{i|j}^{(1L)}$&$\dfrac{|l_i \bigcap l_j|}{|l_j|}+\dfrac{|d_i \bigcap d_j|}{|d_j|}$\\
$K_{i|j}^{(1F)}$&$\dfrac{|l_i \bigcap l_j|}{|l_j|}+\dfrac{|d_i \bigcap d_j|}{|d_i|}$\\
$K_{i|j}^{(2L)}$&$\dfrac{|l_i \bigcap l_j|}{|l_j|}-\dfrac{|d_i \bigcap l_j|}{|l_j|}$\\
$K_{i|j}^{(2F)}$&$\dfrac{|l_i \bigcap l_j|}{|l_j|}-\dfrac{|d_i \bigcap l_j|}{|d_i|}$\\
$K_{i|j}^{(3L)}$&$\dfrac{|l_i \bigcap l_j|}{|l_j|}-\dfrac{|l_i \bigcap d_j|}{|d_j|}$\\
$K_{i|j}^{(3F)}$&$\dfrac{|l_i \bigcap l_j|}{|l_j|}-\dfrac{|l_i \bigcap d_j|}{|l_i|}$\\
\bottomrule
 \end{tabular}
\end{table}

\subsection{Agent-Based Validation of the Model}\label{sec.ag-bas}

For simulating the model, we use the following agent-based framework.

\emph{Users' activity and News' diffusion.} In each time step of the simulation, a given user is active with probability $p_A$. 
When active, a user reads the top $R$ news from her recommendation list, immediately forwarding the ones she likes to her followers; 
in addition, with probability $p_S$ she submits a new piece of news. Users have a limited receiving capability: 
if the number of recommended news becomes greater than the maximum users' stack length $S$, the news with lowest score is removed from the recommendation list. 

\emph{Users' network and Links' rewiring.} 
At any moment, the leaders of each user $j$ should be those $L$ users who have the highest taste similarity with $j$. 
Lacking any prior information, we assume random initial assignment of leaders. As the system gathers more evaluations from users, 
the quality of leaders can be assessed through the similarity metric, and links can be rewired (with the described hybrid strategy) 
in order to connect users to their best leaders. Rewiring of the links takes place after every $u$ time steps.

\emph{Users' tastes and News' attributes.} To model users' judgments of read news we use a vector model where tastes of user $i$ are represented by a $D$-dimensional taste vector 
$\vec{t}_i=(t_i^1,\dots,t_i^D)$ and attributes of news $\alpha$ are represented by a $D$-dimensional attribute vector $\vec{a}_\alpha=(a_\alpha^1,\dots,a_\alpha^D$). 
Similar vector models are often used in semantic approaches to recommendation \cite{CantadorAHAWBS2008}. 
Opinion of user $i$ about news $\alpha$ is based on the overlap of the user's tastes and the news's attributes, which can be expressed by the scalar product 
\begin{equation}\label{eq.ovlp}
\Omega_{i\alpha}=\vec{t}_i\cdot\vec{a}_\alpha
\end{equation}
We assume that user $i$ approves news $\alpha$ ($e_{i\alpha}=+1$) only when $\Omega_{i\alpha}\ge\Delta$; the news is disapproved otherwise ($e_{i\alpha}=-1$). 
Here $\Delta$ is the users' approval threshold: the higher it is, the more demanding the users are. 
To set the taste/attribute vectors, we use two different approaches.
\begin{itemize}
 \item \emph{Homogeneous setting}. Each user has preference for $D_A$ out of $D$ available tastes---that means, each taste vector has $D_A$ elements equal to one (active tastes) 
and the remaining $D-D_A$ elements equal to zero. There are $\binom{D}{D_A}$ such vectors and hence we set the number of users in our system to $U=\binom{D}{D_A}$, 
all having mutually different taste vectors. 
Note that the taste vectors of two users differ at least in two elements, we thus define as \emph{taste-mates} users with exactly two different taste vector elements. 
We also assume that the attributes of a created news $\alpha$ are identical with the tastes of the user $i$ who submits it ($\vec{a}_\alpha\equiv\vec{t}_i$).
 \item \emph{Heterogeneous setting}. Each user has preference for a variable number of the $D$ available tastes---each vector has a different number of elements equal to one (active tastes) 
and the remaining elements are zero. Here users differ not only by their particular tastes, but also by the scope of their interests---which, for a user $i$, is defined as 
$h_i:=\sum_{k=1}^Dt_i^k$ and represents the number of her active tastes, i.e., the area covered by her interests.\footnote{$h_i\equiv D_A$ $\forall i$ in the homogeneous setting: 
each user has the same number of active tastes. To better illustrate the difference between \emph{taste} (or \emph{interest}) and {scope of interests} (or, simply, \emph{scope}), 
consider user $i$ who has preference for news about politics and economics, and user $j$ who instead prefers news about economics, technology and science. 
Then the tastes of $i$ are exactly politics and economics, and we can say that $i$ and $j$ have the same taste for economics (they are both interested in it) and, for instance, for sport (they are both not interested in it). 
The scope of user $i$ is instead equal to two, as she has preference for two kind of news), and the scope of $j$ is three. We can thus say that $i$ has smaller scope than $j$, and that $j$ has bigger scope than $i$.}
There are $2^D$ possible taste vectors and hence in principle we would set $U=2^D$ 
to have all mutually different taste vectors, with taste-mates now defined as users having only one different vector element. 
However in this scenario we are also considering users who like or dislike everything (i.e., the users with taste vectors of all ones or zeros, respectively). 
Hence in order to be more consistent and closer to life, we only consider the users whose taste vectors have a scope $D_A\le h\le D-D_A$. 
This reduces the effective number of users to $U=2^D-2\,\sum_{d=0}^{D_A-1}\binom{D}{d}$.
We finally assume that each news' attribute vector has a fixed number $D_A$ of active attributes (number of ones), 
which are randomly chosen among the active tastes of the user who submits it.
\end{itemize}
Parameters values used in all following simulations are listed in Table \ref{tab.para}. For a discussion on how these parameters may affect simulation results, see \cite{CiminiPRE2012}.
\begin{table}\small
\centering
 \caption{\scriptsize{List of parameters used in simulations.}\label{tab.para}}
 \begin{tabular}{llr}
				& symbol	& value	\\
\toprule %
\emph{Homogeneous setting:}	&		&	\\
Number of users			& $U$		& 3003	\\
Dimension of taste vectors	& $D$		& 14	\\
Number of active tastes 	& $D_A$		& 6	\\
Users' approval threshold	& $\Delta$	& 4	\\
\midrule
\emph{Heterogeneous setting:}	&		&	\\
Number of users			& $U$		& 3498	\\
Dimension of taste vectors	& $D$		& 12	\\
Minimum number of active tastes & $D_A$		& 4	\\
Users' approval threshold	& $\Delta$	& 3	\\
\midrule
\emph{Common parameters:}	&		&	\\
Number of leaders per user	& $L$		& 10	\\
Probability of being active	& $p_A$		& 0.05	\\
Probability of submitting a news& $p_S$		& 0.02	\\
Number of news read when active	& $R$		& 3	\\
Length of recommendation list	& $S$		& 50	\\
Period of the rewiring		& $u$		& 10	\\
Damping factor for time decay	& $\tau$	& 10	\\
Initial value of the similarity	& $s_0$		& $10^{-7}$\\
\bottomrule
 \end{tabular}
\end{table}

\subsection{Measures of the System's Properties}

We measure a number of quantities to describe the leader-follower network and the performance of the social recommender system:
\begin{itemize}
 \item \emph{Leadership Structure}, measured by the cumulative distribution $F(k)=P(f>k)$, where $f$ is the number of a user' followers.
 \item Number of \emph{dead ends} ($de$) in the network, i.e. of users with no followers ($f=0$). 
We are interested in this quantity as, by definition, the news which are uploaded or liked by dead ends cannot spread to any other users, 
hence having many dead ends leads to a waste of user input and significantly reduces the diversity of news propagating over the network.
 \item \emph{Link reciprocity}, the tendency of node pairs to form connections between each other. 
It is defined as the ratio of the number of bi-directed links to the total number of links in the network:
$$r:=\frac{\sum_{ij}a_{ij}a_{ji}}{\sum_{ij}a_{ij}}$$
 \item \emph{Average differences}. In a computer simulation we have the luxury of knowing users' taste vectors and hence we can compute in how many elements 
the taste vector of a user differs from the taste vector of her leader. By averaging over all users and all their leaders we obtain the average number of differences 
$ad:=\tfrac{1}{UL}\sum_i\sum_{l\in L_i}\|\vec{t}_i-\vec{t}_l\|_1$. The fewer are the differences, the better is the assignment of leaders and 
the better the network has adapted to users' tastes.
 \item \emph{Recommendation efficiency.} When making recommendations, it is possible to fall into two different kinds of error: 
recommending content that users does not like, and not recommending content that users does like. These errors are known respectively as \emph{false positives} (FP) 
and \emph{false negatives} (FN). To complete the picture, \emph{true positives} (TP) are recommendations of content that users does like, 
and \emph{true negatives} (TN) are lacks of recommendation of content that user does not like \cite{Neyman-Biometrika-1928}.
These quantities can be used to assess the quality of recommendations by measuring:
\begin{enumerate}
     \item \emph{Precision} $p:=\frac{\mbox{TP}}{\mbox{TP+FP}}$
     \item \emph{Recall} $1-\beta:=\frac{\mbox{TP}}{\mbox{TP+FN}}$
\end{enumerate}
It is important to remark that in our model these quantities strongly depend on $L$, the number of leaders per user: 
big values of $L$ imply more recommendations, which can have the effect of decreasing $p$ and increasing $1-\beta$, and vice versa.
\end{itemize}

\section{Results of agent-based simulations}\label{sec.result}

\begin{figure}
   \centering
   \includegraphics[width=\textwidth]{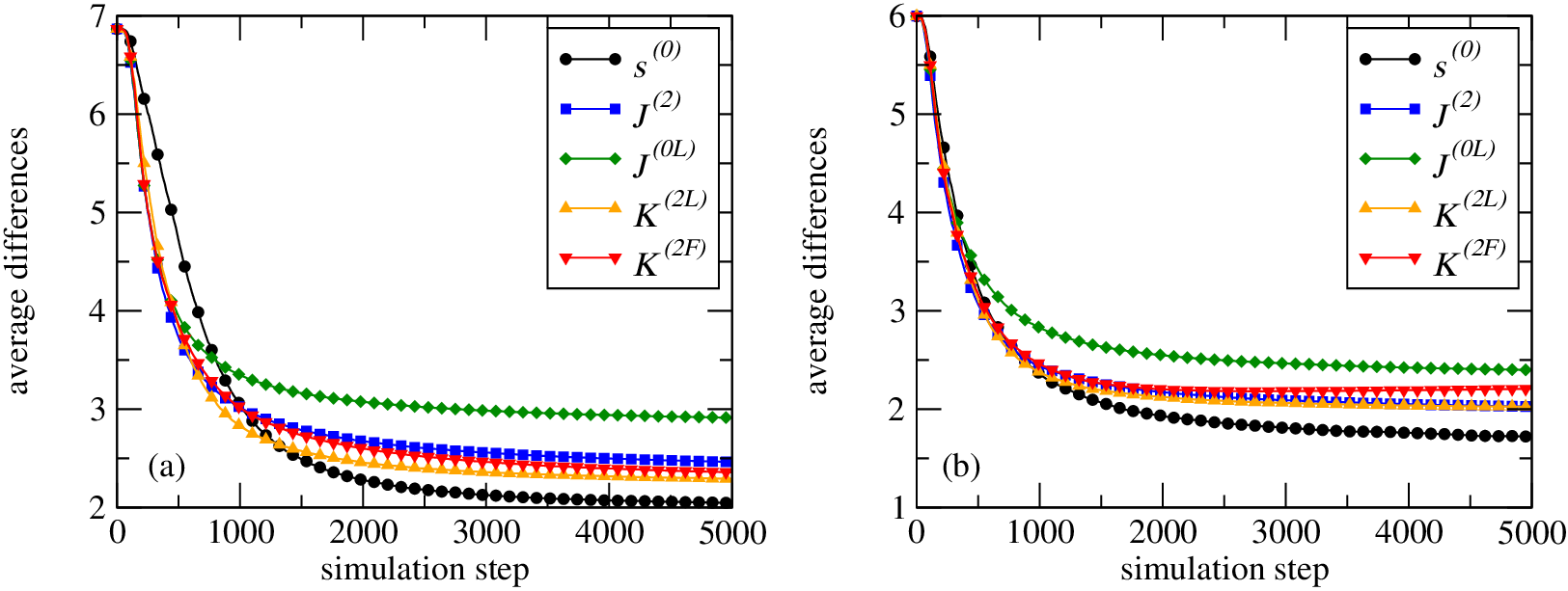}
   \caption{Evolution of the average differences in the system for the homogeneous (a) and the heterogeneous (b) settings, 
and for some representative definitions of the similarity measure.}\label{fig.evol}
\end{figure}

In this section we present the results of agent-based simulations of the social recommendation model.
We first check if the system is able to improve the leader-follower network from its initial random configuration. 
Figure \ref{fig.evol} shows the average differences at different times steps of the network's evolution for both homogeneous and heterogeneous settings. 
For readability reasons, we only report the evolution of the system under some of the definitions of the similarity measure 
(those, as will be discussed later, that are best performing). We observe that the average differences are gradually improved in the system, 
meaning that users are able to find better leaders---whatever the definition of the similarity. After an equilibration period, the system converges to a stationary network configuration 
(whose properties are reported in table \ref{tab.result}). The quality of the equilibrium state can be assessed as follows:
\begin{itemize}
 \item In the homogeneous setting, taste-mates differ in exactly two taste vector elements: the lowest possible value of the average differences is two, 
which corresponds to a globally optimal assignment of leaders in the network. Hence the quality of the final state can be estimated by how close the average differences are to two. 
Note that as for each user there are $N=D_A(D-D_A)$ taste-mates, the optimal state is unique only if $L=N$. If $L<N$, there are different possible optimal states, 
which are determined by initial conditions and users' dynamics. If $L>N$ (which however is not our case), the average differences are more than two even in the optimal state.
 \item In the heterogeneous setting, taste-mates differ instead in only one taste vector element, and the lowest possible value of the average differences is hence one. 
In a situation where $U=2^D$, there would be $D$ taste-mates for each user, hence the optimal state would be unique if $L=D$, whereas if $L<D$ there would be different optimal states, 
and finally if $L>D$ the average differences would be greater than one even in the optimal state. 
However we are considering here only users with $D_A\le h\le D-D_A$, hence users with $h=D_A$ or $h=D-D_A$ only have $D-D_A$ taste-mates. 
Since we have set $D-D_A<L<D$, we do not expect the average differences to ever reach one. 
In addition, we notice that here the number of taste-mates for each user is smaller compared to the homogeneous setting, hence getting close to the optimal state is more difficult. 
Finally, whereas in the homogeneous case the notion of taste-mates is simple, here we still have to investigate whether a good leader should have more or less active tastes 
than her followers---that would again reduce the number of available good leaders for each user. 
\end{itemize}
Figure \ref{fig.evol} shows that in both settings, with $s^{(0)}$ the system reaches the lowest average differences. 
With $J^{(0L)}$, which only considers likes, one obtains the poorest performance---but the stationary state is not far from the optimal one. 
Using additional terms as in $K^{(2L)}$ and $K^{(2F)}$ brings the system very close to the final state of $s^{(0)}$. 
The stationary values of the average differences for these and other similarities are reported in table \ref{tab.result}. 
\begin{figure}
   \centering
   \includegraphics[width=\textwidth]{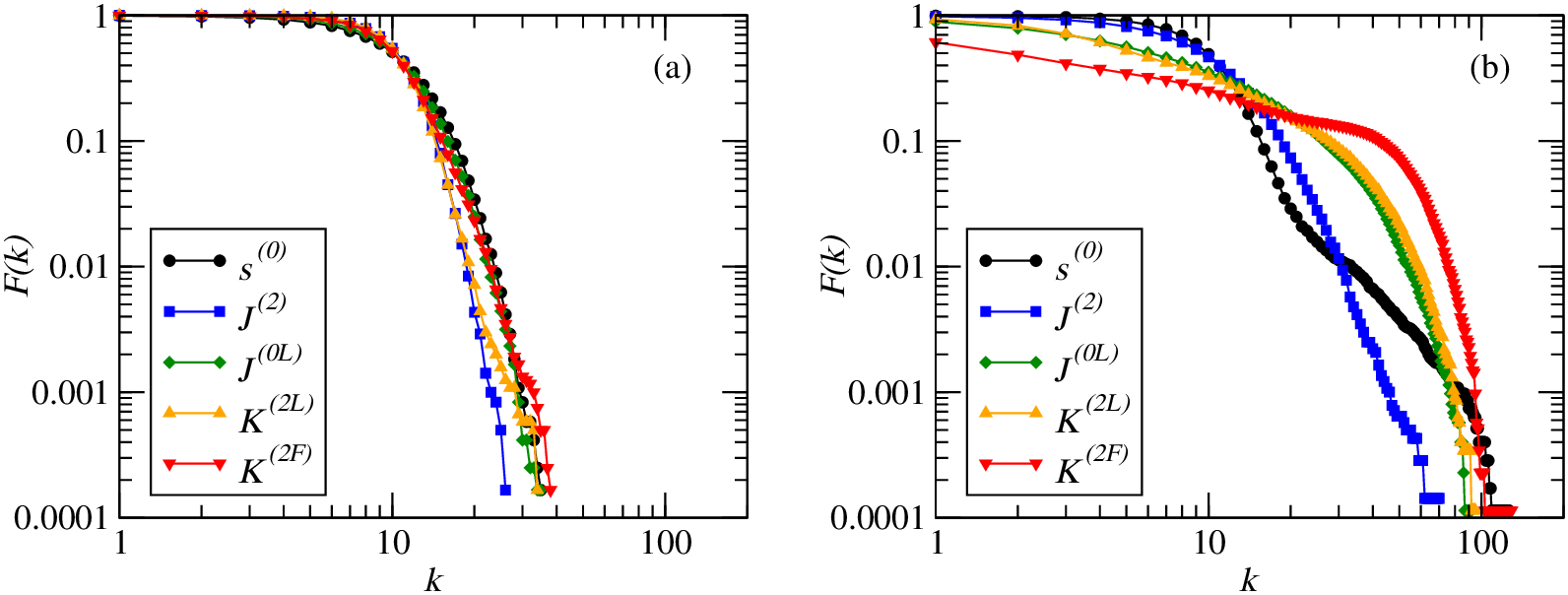}
   \caption{Cumulative distribution functions of the number of users' followers at equilibrium for the homogeneous (a) and the heterogeneous (b) setting, 
and for some representative definitions of the similarity measure.}\label{fig.degree}
\end{figure}
\begin{table}
\centering
 \caption{Properties of the network at equilibrium and performance of the recommender system (results are averaged over independent realizations). 
Values of $de$, $p$ and $1-\beta$ are percentages. Note that the precision values in the two settings have as reference values the precisions of recommending at random, 
which are 15.6\% and 29.6\% in the homogeneous and heterogeneous setting, respectively. \label{tab.result}}
 \begin{tabular}{lcccccccccccc}
\toprule %
		&&	\multicolumn{5}{c}{\emph{homogeneous setting}}				& &	\multicolumn{5}{c}{\emph{heterogeneous setting}}			\\
		&&	$ad$	&	$r$	&	$de$	&	$p$ 	&	$1-\beta$& &	$ad$	&	$r$	&	$de$	&	$p$	&	$1-\beta$\\
\toprule %
\rowcolor[gray]{.9}
$s^{(0)}$	&&	2.05	&	0.67	&	0.5	&	45.8	&	36.9	& &	1.72	&	0.63	&	0.6	&	61.5	&	21.9	\\
$J^{(0)}$	&&	3.08	&	0.73	&	0.3	&	39.0	&	16.8	& &	2.73	&	0.67	&	1.1	&	50.2	&	10.3	\\
\midrule
$J^{(1)}$	&&	3.33	&	0.71	&	0.7	&	35.9	&	10.9	& &	2.81	&	0.66	&	1.2	&	47.8	&	9.5	\\
\rowcolor[gray]{.9}
$J^{(2)}$	&&	2.46	&	0.80	&	0.0	&	46.0	&	29.4	& &	2.03	&	0.64	&	1.6	&	59.2	&	14.9	\\
$J^{(3)}$	&&	2.53	&	0.49	&	4.3	&	41.7	&	20.4	& &	2.09	&	0.49	&	6.9	&	50.4	&	11.6	\\
\midrule
$s^{(0L)}$	&&	4.61	&	0.59	&	5.5	&	28.6	&	4.6	& &	3.19	&	0.62	&	2.4	&	47.8	&	9.0	\\
$s^{(0F)}$	&&	3.76	&	0.66	&	1.3	&	30.9	&	7.3	& &	2.62	&	0.60	&	2.8	&	51.1	&	10.1	\\
\rowcolor[gray]{.9}
$J^{(0L)}$	&&	2.91	&	0.76	&	0.2	&	42.3	&	22.5	& &	2.40	&	0.46	&	10.8	&	59.8	&	12.5	\\
$J^{(0F)}$	&&	3.05	&	0.24	&	35.6	&	35.6	&	6.2	& &	2.55	&	0.10	&	62.3	&	35.5	&	5.8	\\
\midrule
$K^{(1L)}$	&&	3.12	&	0.71	&	0.5	&	36.5	&	14.3	& &	2.55	&	0.59	&	2.1	&	52.4	&	12.8	\\
$K^{(1F)}$	&&	3.20	&	0.68	&	1.7	&	38.0	&	16.9	& &	2.46	&	0.30	&	33.5	&	59.0	&	7.6	\\
\rowcolor[gray]{.9}
$K^{(2L)}$	&&	2.29	&	0.81	&	0.1	&	47.7	&	33.9	& &	2.02	&	0.47	&	7.0	&	66.3	&	19.3	\\
\rowcolor[gray]{.9}
$K^{(2F)}$	&&	2.36	&	0.77	&	0.4	&	46.8	&	32.9	& &	2.21	&	0.17	&	38.9	&	66.7	&	11.3	\\
$K^{(3L)}$	&&	2.43	&	0.53	&	0.9	&	44.0	&	25.7	& &	2.01	&	0.54	&	1.3	&	59.9	&	17.8	\\
$K^{(3F)}$	&&	2.44	&	0.45	&	5.4	&	42.9	&	21.9	& &	2.07	&	0.60	&	3.6	&	56.0	&	14.7	\\
\bottomrule
 \end{tabular}
\end{table}
\newline\newline
Moving further, we investigate the leadership structure (figure \ref{fig.degree}) and the properties of the network at equilibrium, as well as the performance of the recommender system 
(table \ref{tab.result}). As it was natural to expect, in the homogeneous setting the stationary degree distribution is rather homogeneous (with very few dead ends) 
and the reciprocity is high, whatever the definition of the similarity. The only exception is $J^{(0F)}$: here users who happen to have few likes are not favored as leaders 
($l_i\bigcap l_j$ is not necessarily small compared to $l_j$, but it is small compared to an average $l_i$), so they start to lose followers and the news they forward reach only a few users; 
this in turns causes $l_i\bigcap l_j$ to remain small while $l_i$ grows, so they again lose attractiveness, and so forth. As a results, these users end up with no followers. 
About the recommender system's performance, using $s^{(0)}$ (the best for average differences) yields the highest recall, 
whereas using $K^{(2L)}$ and $K^{(2F)}$ the highest precision. Moreover, all similarities highlighted in the table are better than the others both in precision and recall. 
\newline
Things become more interesting in the heterogeneous setting. Here the stationary degree distribution has a longer tail (as it happens in many real systems) 
and the reciprocity is generally lower with respect to the homogeneous case. $s^{(0)}$ is still the best for average differences and recall, 
and the resulting network is again rather homogeneous (low $de$ and high $r$), whereas 
$K^{(2L)}$ and $K^{(2F)}$ have still the best precision (the gap with respect to $s^{(0)}$ increases here). However, for $K^{(2F)}$ the recall drops significantly, 
and this is caused by high $de$ ratio. In this case, users with many likes are not favored ($d_i\bigcap l_j$ is usually smaller than $l_j$, 
but can be as big as an average $d_i$ causing a $-1$ contribution to the similarity) and end up with no followers.\footnote{This feature only appears in the heterogeneous setting 
which consists of users who by construction like more news than the others. In the homogeneous setting, a similar situation may arise only because of fluctuations 
which are however tamed by the first term $|l_i\bigcap l_j|/|l_j|$.} The same happens for $K^{(1F)}$ (for users with few dislikes, $d_i\bigcap d_j$ is not necessary small compared to $d_j$, 
but it is small compared to an average $d_i$) and again for $J^{(0F)}$. 
\newline
Note that among the basic asymmetric indices, $J^{(0L)}$ performs remarkably better than $s^{(0L)}$ and $s^{(0F)}$---despite using only likes, 
and as expected it is better than $J^{(0F)}$. This justifies using $J^{(0L)}$ for building 
the set of $K$ similarities, with the result of $K^{(2L)}$ and $K^{(2F)}$ further increasing precision and recall in the system. 
Again, we can explain why these two metrics outstand the other $K$s: $K^{(2L)}$ and $K^{(2F)}$ are the only to feature in the second term the set $l_j$, 
which is the most informative to consider as it contains all the news $j$ passes to $i$---which determine $i$'s satisfaction, and hence the quality of the recommendation process. 
\begin{table}
\centering
 \caption{Pearson correlation coefficients $c_{(\Delta t,s)}$ and $c_{(\Delta h,s)}$ between $\Delta t$ or $\Delta h$ and $s$ for leader-follower pairs in equilibrium networks. 
The left column refers to the homogeneous setting (where $\Delta h_{i|j}=0$ $\forall i,j$), the other two columns to the heterogeneous one.\label{tab.simil}}
 \begin{tabular}{lccccc}
\toprule %
		&	&$c_{(\Delta t,s)}$&	&$c_{(\Delta t,s)}$&$c_{(\Delta h,s)}$\\
\toprule %
\rowcolor[gray]{.9}
$s^{(0)}$	&	&	-0.68	&	&	-0.30	&	0.00	\\
$J^{(0)}$	&	&	-0.29	&	&	-0.26	&	-0.01	\\
\midrule
$J^{(1)}$	&	&	-0.17	&	&	-0.21	&	-0.00	\\
\rowcolor[gray]{.9}
$J^{(2)}$	&	&	-0.40	&	&	-0.33	&	-0.13	\\
$J^{(3)}$	&	&	-0.36	&	&	-0.38	&	0.20	\\
\midrule
$s^{(0L)}$	&	&	-0.05	&	&	-0.19	&	-0.05	\\
$s^{(0F)}$	&	&	-0.13	&	&	-0.25	&	0.03	\\
\rowcolor[gray]{.9}
$J^{(0L)}$	&	&	-0.36	&	&	-0.39	&	-0.14	\\
$J^{(0F)}$	&	&	-0.37	&	&	-0.44	&	0.35	\\
\midrule
$K^{(1L)}$	&	&	-0.20	&	&	-0.28	&	0.03	\\
$K^{(1F)}$	&	&	-0.24	&	&	-0.36	&	-0.31	\\
\rowcolor[gray]{.9}
$K^{(2L)}$	&	&	-0.46	&	&	-0.42	&	-0.33	\\
\rowcolor[gray]{.9}
$K^{(2F)}$	&	&	-0.47	&	&	-0.45	&	-0.51	\\
$K^{(3L)}$	&	&	-0.42	&	&	-0.42	&	0.05	\\
$K^{(3F)}$	&	&	-0.40	&	&	-0.34	&	0.02	\\
\bottomrule
 \end{tabular}
\end{table}
\newline\newline
We now investigate further the role of various users in the heterogeneous setting. In particular, we consider each leader-follower pair in the system 
and define their taste-vector difference $\Delta t_{ij}=\|\vec{t}_i-\vec{t}_j\|$ and their scope difference $\Delta h_{i|j}=h_j-h_i$. 
We then study which pairs are favored by the used similarity measure by computing the Pearson correlation coefficients $c_{(\Delta t,s)}$ 
(between $\Delta t$ and $s$) and $c_{(\Delta h,s)}$ (between $\Delta h$ and $s$). Table \ref{tab.simil} shows these values, together with 
$c_{(\Delta t,s)}$ for the homogeneous setting as a benchmark. 
\newline
As expected, in the homogeneous setting all $c_{(\Delta t,s)}$ are negative: $s$ and $\Delta t$ are anticorrelated, i.e., the smaller the difference, the higher the similarity. 
$s^{(0)}$ yields the highest $c$ in absolute value, followed by $K^{(2L)}$ and $K^{(2F)}$. 
We see that the ability to distinguish users' pairs with different $\Delta t$ is directly related to final average differences and system's performance (table \ref{tab.result}). 
\newline
In the heterogeneous setting, $c_{(\Delta t,s)}$ values are still negative and very close to the ones of the previous setting---with the exception of $s^{(0)}$ whose $c$ is halved.
With respect to $c_{(\Delta h,s)}$, as expected, symmetric measures cannot distinguish $\Delta h_{i|j}$ from $\Delta h_{j|i}$, hence we obtain no correlation in this case. 
$J^{(0F)}$ favors instead users with higher $h$ (more likes, as explained before). For other metrics, we generally observe a weak correlation, with some important exceptions: 
$J^{(2)}$, $J^{(0L)}$, $K^{(1F)}$, $K^{(2L)}$ and $K^{(2F)}$, ordered by increasing anticorrelation. These metrics are able to distinguish user pairs with different scopes 
and favor users with smaller scope as leaders, which then leads to superior recommendations (table \ref{tab.result}). 
\newline
To show that this is the case, consider a user pair with $\Delta t_{ij}$ small and $h_i>h_j\Rightarrow\Delta h_{i|j}<0$ : users $i$ and $j$ have very similar tastes, 
but $j$ is slightly more demanding than $i$. Since $j$ is $i$'s leader, in the news spreading process $j$ blocks all the news she dislikes so that $i$ will like 
almost all the news that she gets from $j$. This means that $j$ is a good leader for $i$ as she forwards only the news matching $i$'s tastes. 
If we do not consider missing evaluations, this situation can be roughly described as 
$|l_i \bigcap l_j|\simeq|l_j|$, $|d_i \bigcap l_j|\simeq0$, $|l_i \bigcap d_j|\neq0$, $|d_i \bigcap d_j|\simeq|d_i|<|d_j|$. 
The opposite case $h_i<h_j$ is instead not favorable for $i$ as in this case her leader $j$ could also forward news that $i$ would not like (as $j$ is less demanding than $i$). 
Here $|l_i \bigcap l_j|\simeq|l_i|<|l_j|$, $|d_i \bigcap l_j|\neq0$, $|l_i \bigcap d_j|\simeq0$, $|d_i \bigcap d_j|\simeq|d_j|<|d_i|$. 
The situation is not symmetric and that is why the reciprocity coefficient in the heterogeneous setting drops. 
Given these two cases, $J^{(0L)}$ favors the first one as by definition, $J^{(0L)}_{i|j}\simeq1$ if $h_i>h_j$ but $J^{(0L)}_{j|i}\equiv J^{(0F)}_{i|j}<1$ if $h_i<h_j$. 
The other $K$ metrics are built from $J^{(0L)}$ by introducing additional terms. These terms for $K^{(0L)}$, $K^{(0F)}$ and, in lesser degree,\footnote{We already discussed 
that the term $|d_i \bigcap d_j|$ is not the most reliable for computing similarity.} $K^{(1F)}$
promote the first case (and hence users' satisfaction) even more, whereas the terms for $K^{(1L)}$, $K^{(3L)}$ and $K^{(3F)}$ go in the opposite direction: 
the corresponding overall $c_{(\Delta h,s)}$ becomes close to zero and users' satisfaction lowers significantly. 
The system's behavior hence strongly depends on the particular similarity definition used.
\begin{table}
\centering
 \caption{Pearson correlation coefficients of $h$ with $f$, $p$ and $1-\beta$ in the heterogeneous setting.\label{tab.scope}}
 \begin{tabular}{lcccc}
\toprule %
		&	&$c_{h,f}$	&$c_{h,p}$	&$c_{h,1-\beta}$\\
\toprule %
\rowcolor[gray]{.9}
$s^{(0)}$	&	&	0.18	&	0.96	&	-0.33	\\
$J^{(0)}$	&	&	-0.26	&	0.89	&	-0.72	\\
\midrule
$J^{(1)}$	&	&	-0.27	&	0.89	&	-0.54	\\
\rowcolor[gray]{.9}
$J^{(2)}$	&	&	-0.62	&	0.94	&	-0.84	\\
$J^{(3)}$	&	&	0.47	&	0.92	&	-0.80	\\
\midrule
$s^{(0L)}$	&	&	0.04	&	0.93	&	0.15	\\
$s^{(0F)}$	&	&	0.06	&	0.93	&	-0.01	\\
\rowcolor[gray]{.9}
$J^{(0L)}$	&	&	-0.74	&	0.91	&	-0.85	\\
$J^{(0F)}$	&	&	0.49	&	0.96	&	-0.48	\\
\midrule
$K^{(1L)}$	&	&	-0.56	&	0.92	&	-0.62	\\
$K^{(1F)}$	&	&	-0.74	&	0.92	&	-0.75	\\
\rowcolor[gray]{.9}
$K^{(2L)}$	&	&	-0.74	&	0.92	&	-0.93	\\
\rowcolor[gray]{.9}
$K^{(2F)}$	&	&	-0.73	&	0.91	&	-0.89	\\
$K^{(3L)}$	&	&	-0.15	&	0.93	&	-0.88	\\
$K^{(3F)}$	&	&	0.12	&	0.92	&	-0.86	\\
\bottomrule
 \end{tabular}
\end{table}
\begin{figure}
   \centering
   \includegraphics[width=\textwidth]{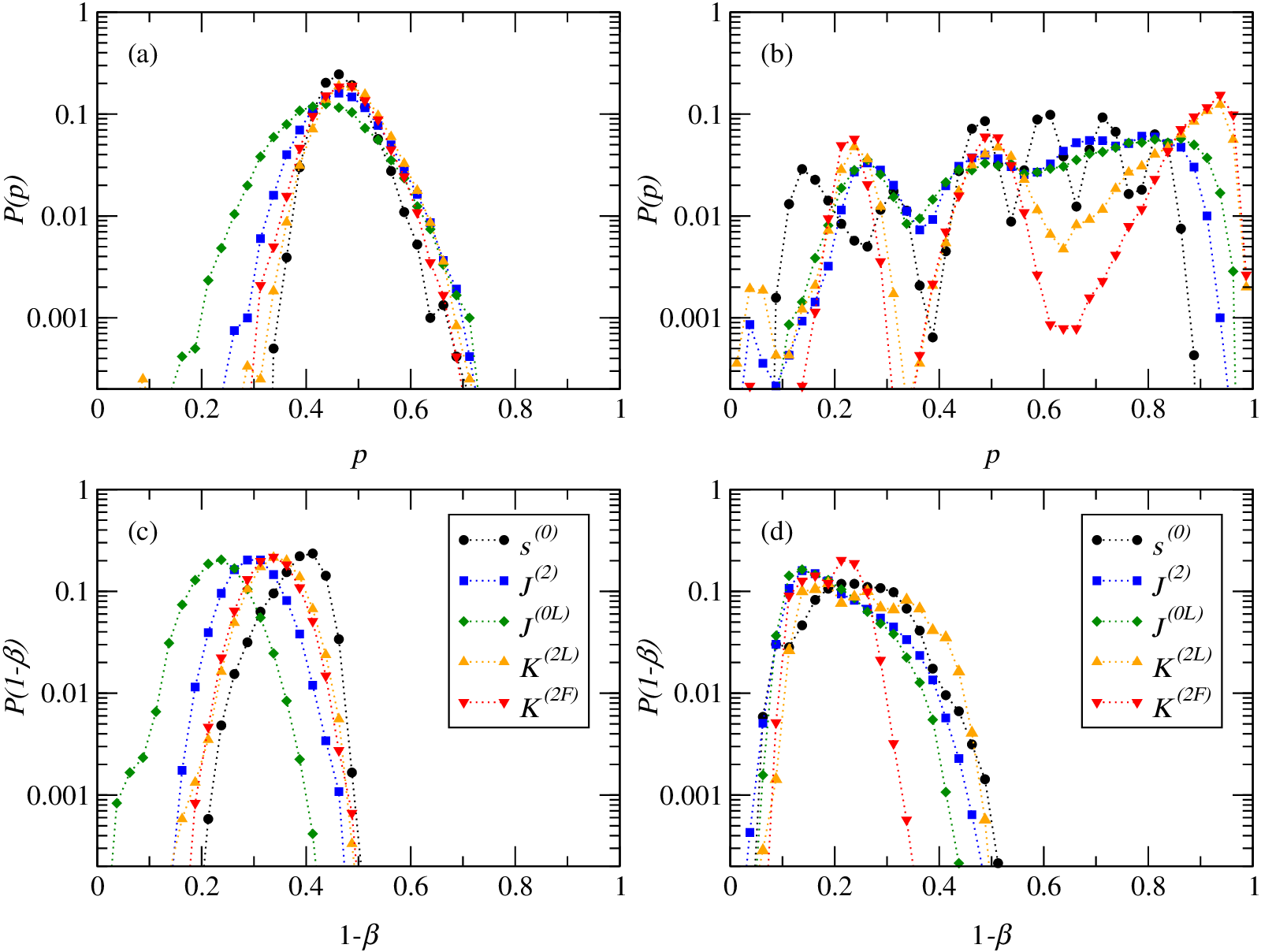}
   \caption{Probability distributions for individual users' precision (upper panels) and recall (lower panels) values 
for the homogeneous [(a),(c)] and the heterogeneous [(b),(d)] settings.}\label{fig.histo}
\end{figure}
\newline\newline
In addition we study how individual users' features depend on the scopes of interests $h$. Table \ref{tab.scope} reports the Pearson correlation coefficients of $h$ 
with $f$ (number of users' followers), $p$ and $1-\beta$ (precision/recall of the recommendations for the user).
Concerning the number of followers, users with small $h$ (more demanding) become popular leaders when using $J^{(2)}$, $J^{(0L)}$, $K^{(1F)}$, $K^{(2L)}$ and $K^{(2F)}$, 
as expected from the previous discussion. This feature is still present when using $K^{(1L)}$ and $K^{(3L)}$, 
as in these cases the effect of the damping terms (the ones in parenthesis) is to favor users with many dislikes, i.e. with small $h$. 
For $s^{(0)}$ instead the correlation is small but positive: users with high $h$ are slightly favored as their bad filtering ability is not penalized enough by a symmetric measure, 
but on the contrary it is compensated by the big number and variety of news they forward (which increases the overlap of rating histories with any other user, 
and hence the probability of being chosen as a leader by that user). 
Finally, for $J^{(0F)}$ (and $J^{(3)}$ which features $l_i$ in the denominator of the second term) 
users with small $h$ are without followers---as a result of the previous discussion.
Moving further, naturally precision is always better for users with high $h$, who are less demanding. Recall instead has the opposite trend: 
demanding users with small $h$ are potentially interested in a few news, which they easily get from well-chosen leaders, 
whereas users with big $h$ would be interested in many more news, but still have the same finite receiving capability. 
\newline\newline
Finally we plot in figure \ref{fig.histo} the probability distributions of precision and recall values for individual users. 
We observe that in the homogeneous setting these distribution have a well defined peak, whereas in the heterogeneous setting $p$ and $1-\beta$ values vary more among users. 
This represents a natural (and expected) outcome of the particular way to model users' tastes.

\section{Testing on real data}

In this section, we test the similarity measures presented in Sec. \ref{sec.similarity} on real data. 
To this end, we employ users' assessments from two online rating platforms to build two corresponding leader-follower networks for social recommendation 
(in the spirit of the model presented in this work), and examine if the results obtained by artificial simulations can be recovered from real users' preferences. 

The dataset that we consider are from:
\begin{itemize}
 \item MovieLens,\footnote{http://www.movielens.org} a movie recommendation website in which users can rate movies to get personalized recommendations. 
The dataset that we analyze is freely available at www.grouplens.org, and consists of over 1,000,000 anonymous ratings on approximately 3,900 movies made by about 6,000 MovieLens users 
since year 2000. The rating system is integer, with scale from 1 (worst) to 5 (best).
 \item Netflix,\footnote{http://www.netflix.com} the world's largest online movie rental service where users can rate the movies they watch. 
The dataset that we analyze is freely provided for the Netflix Prize (http://www.netflixprize.com), and consists of over 100,000,000 ratings from about 480,000 anonymous Netflix customers 
on 17,000 movie titles. Again, ratings are integers from 1 to 5. 
\end{itemize}
For each dataset, and for each similarity measure, we use the ratings given by users to build a leader-follower network: 
each user selects as leaders $L=10$ other users which are the most similar to her.
Since the similarity definitions in Sec. \ref{sec.similarity} are based on binary assessments (like/dislike), in order to restore this setting 
we regard ratings 4 and 5 as likes and ratings 1, 2, 3 as dislikes. In this way, we have almost the same number of likes and dislikes for each dataset. 
Note also that the process of selecting for each user the most similar leaders is rather demanding, requiring approximately $U\times M$ operations (here $M$ is the total number of ratings). 
We hence restrict our analysis to subsets of 1,000 randomly selected users. For Netflix, we also consider only 4,000 randomly selected movies (and respective ratings). 
We then average our results over various independent selections of the subsets. 
\newline
After the social network of users is constructed, movie recommendations for a given user are directly given as the movies liked by her leaders 
(we have this information from the real data itself). Note that in order to produce enough recommendations, we do not consider if in reality the user has rated a movie before her leaders, 
or if a movie recommended from a leader actually come from this leader's leaders, and we do not consider damping effects. 
In other words, we use the leader-follower network to have a single-step spreading system, 
in which every movie liked by a user in reality is recommended to her followers. We then obtain precision and recall for these recommendations. 
As before, precision is the ratio of good recommendations (i.e., recommendations for the movies that users liked in reality) to the total number of recommendations for the movies 
liked or disliked by users in reality, whereas recall is the ratio of good recommendations to the total number of movies liked in reality by users. 
\begin{table}
\centering
 \caption{Properties of the artificial leader-follower networks and quality of the recommendations (results are averaged over independent subset selections). 
Values of $de$, $p$ and $1-\beta$ are percentages. The precision values can be compared to the precisions of recommending at random 
(i.e. the percentages of likes among the ratings), which are 61.7\% for MovieLens and 59.8\% for Netflix.\label{tab.emph}}
 \begin{tabular}{lcccccccccc}
\toprule %
		&	&	\multicolumn{4}{c}{\emph{Movielens}}			&	&	\multicolumn{4}{c}{\emph{Netflix}}			\\
		&	&	$r$	&	$de$	&	$p$	&	$1-\beta$&	&	$r$	&	$de$	&	$p$	&	$1-\beta$\\
\toprule %
\rowcolor[gray]{.9}
$s^{(0)}$	&	&	0.25	&	11.7	&	77.7	&	79.8	&	&	0.26	&	20.1	&	76.2	&	82.9	\\
$J^{(0)}$	&	&	0.21	&	27.3	&	70.7	&	52.0	&	&	0.24	&	10.8	&	66.8	&	57.0	\\
\midrule
$J^{(1)}$	&	&	0.22	&	24.2	&	70.3	&	52.4	&	&	0.24	&	14.0	&	65.5	&	57.4	\\
\rowcolor[gray]{.9}
$J^{(2)}$	&	&	0.15	&	18.6	&	71.4	&	49.9	&	&	0.22	&	7.7	&	67.4	&	54.0	\\
$J^{(3)}$	&	&	0.14	&	33.5	&	71.1	&	49.6	&	&	0.21	&	11.5	&	66.2	&	55.6	\\
\midrule
$s^{(0L)}$	&	&	0.22	&	38.3	&	76.2	&	59.9	&	&	0.23	&	26.1	&	82.9	&	61.4	\\
$s^{(0F)}$	&	&	0.18	&	4.0	&	69.0	&	64.2	&	&	0.23	&	3.8	&	63.4	&	71.7	\\
\rowcolor[gray]{.9}
$J^{(0L)}$	&	&	0.22	&	37.8	&	78.4	&	59.0	&	&	0.27	&	22.6	&	82.2	&	69.0	\\
$J^{(0F)}$	&	&	0.13	&	17.0	&	69.0	&	65.1	&	&	0.16	&	5.5	&	62.9	&	72.6	\\
\midrule
$K^{(1L)}$	&	&	0.24	&	29.4	&	74.4	&	59.5	&	&	0.22	&	21.5	&	78.2	&	66.9	\\
$K^{(1F)}$	&	&	0.07	&	30.6	&	69.4	&	81.0	&	&	0.08	&	9.6	&	73.6	&	77.0	\\
\rowcolor[gray]{.9}
$K^{(2L)}$	&	&	0.17	&	40.6	&	88.1	&	55.7	&	&	0.24	&	24.3	&	96.1	&	66.8	\\
\rowcolor[gray]{.9}
$K^{(2F)}$	&	&	0.15	&	43.5	&	86.8	&	54.7	&	&	0.18	&	12.7	&	92.8	&	64.4	\\
$K^{(3L)}$	&	&	0.16	&	41.9	&	78.5	&	57.2	&	&	0.21	&	21.9	&	80.4	&	67.4	\\
$K^{(3F)}$	&	&	0.17	&	44.3	&	79.3	&	57.1	&	&	0.24	&	17.8	&	80.4	&	67.6	\\
\bottomrule
 \end{tabular}
\end{table}
\begin{figure}
   \centering
   \includegraphics[width=\textwidth]{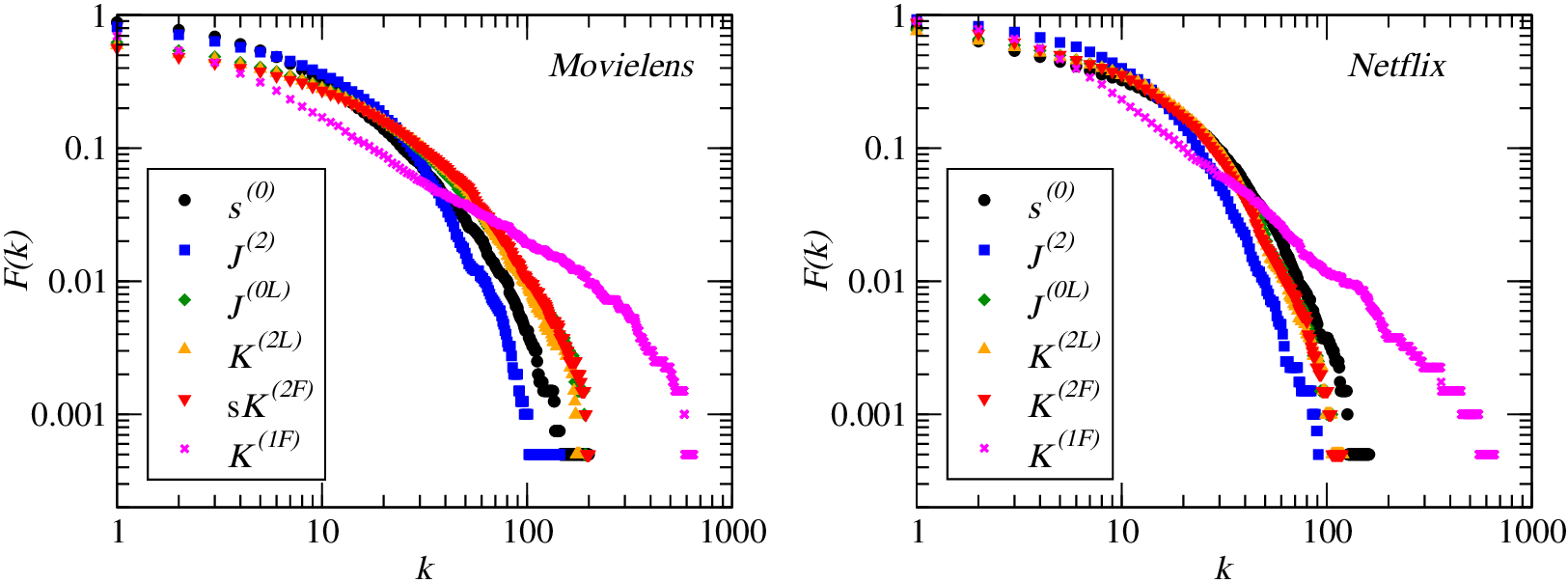}
   \caption{Cumulative distribution function of the number of users' followers for MovieLens and Netflix.}\label{fig.degree_real}
\end{figure}
\newline\newline
Table \ref{tab.emph} reports the properties of the constructed social networks and of the recommendations for the two datasets, 
and Figure \ref{fig.degree_real} shows the leadership structure of the networks. 
We immediately notice that the degree distributions are broad and the reciprocity values are relatively low (compared to the values of Table \ref{tab.result})---even 
with a symmetric measure $s^{(0)}$. As a consequence, there are many dead ends in the network. 
This kind of structure recalls the heterogeneous case more than the homogeneous one (cf. Table \ref{tab.result}). 
Note that the differences between results from agent-based simulations and empirical findings (lower reciprocity and broader degree distribution in the latter case) 
can be attributed to the fact that the real scenario is even more heterogeneous than what the heterogeneous setting is supposed to model: 
real users are diverse not only in their particular tastes and scopes of interest, but also in their behavior, features, social influence, and many other factors. 
For instance, by introducing users' heterogeneity in usage frequency and evaluation noise (as in \cite{CiminiEPJB2011}, but see also \cite{ZhouPLoS2011}), 
one observes a degree distribution which is very broad or even scale-free. 
\newline\newline
As for recommendation metrics, we confirm that also for real data $K^{(2L)}$ and $K^{(2F)}$ are best in precision, whereas $s^{(0)}$ is best for recall. 
The gap in precision between $K^{(2L)}$ and $s^{(0)}$ suggests that asymmetric and heterogeneous tastes better describe users' preferences. 
Finally, for $K^{(1F)}$ we notice a peculiar behavior that we do not observe in simulations: it produces the longest tails of $P(k)$, few dead ends and very low reciprocity. 
The high value of recall obtained in this case can be attributed to this particular network structure. 
\begin{figure}
   \centering
   \includegraphics[width=\textwidth]{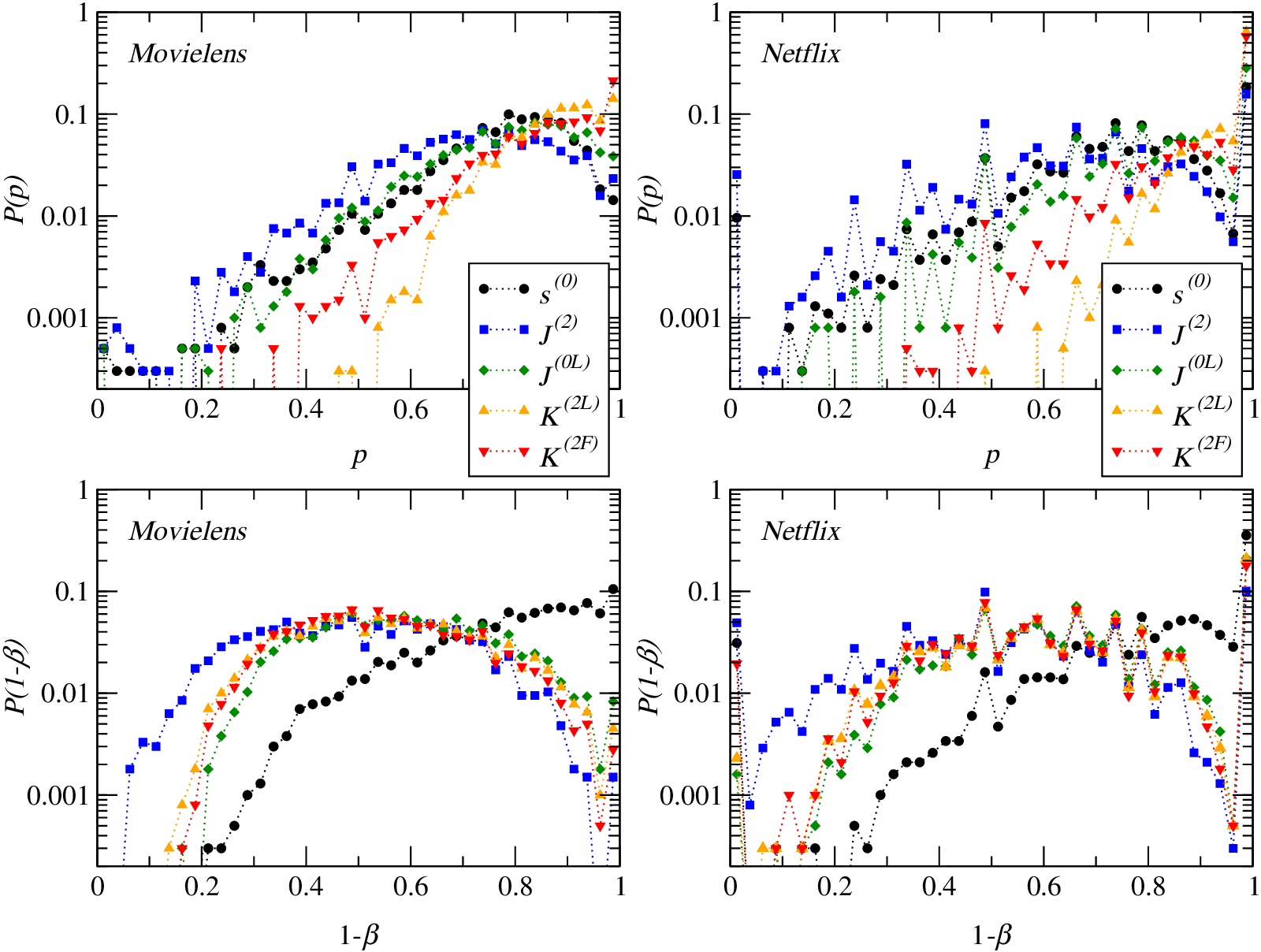}
   \caption{Probability distributions for individual users' precision (upper panels) and recall (lower panels) values for Movielens and Netflix.}\label{fig.histo_real}
\end{figure}
\newline\newline
Finally, we plot in figure \ref{fig.histo_real} the probability distributions of precision and recall values for individual users in the real datasets. 
Both for Movielens and Netflix, we observe that these distributions are broad, suggesting again that the heterogeneous vector setting represents a better way to model real users' tastes.

\section{Conclusion}

In this work, we have studied the role of taste similarity in social networks where social recommendation is at work. 
Building on the adaptive model of social recommendation introduced in \cite{MedoEPL2009}, we introduced two settings for modeling users' tastes, 
and used agent-based modeling to study the evolution and properties of the system under these different settings. 
After cross-checking our results with real data, we concluded that modeling users with different scopes of interests 
produces results closer to those obtained from real user actions.
\newline\newline
Besides modeling users' tastes, we tackled in this work also the complementary issue of quantifying taste similarity from users' actions. 
We investigated several definitions of users' similarity both in agent-based simulations and in real data, 
revealing which definitions are more reliable in measuring taste similarity, and more effective in constructing leader-follower networks for social recommendation. 
In particular, we showed that asymmetric similarities are able to reveal if a user is more or less demanding than another, and generally yield better results. 
In addition, we discussed which sets of actions are more appropriate to consider when building a taste similarity that can enhance both precision and recall of the social recommendation process. 
\newline\newline
We remark that in our model we neglect some aspects of the dynamics of social recommendation and user behavior. 
For instance, real users are not infallible and thus they can commit mistakes in selecting their leaders. At the same time, each user can have her own selection criteria (i.e., her own similarity metric). 
Both these interesting aspects would require extensive consideration, and will be addressed in future works. 
Finally, our model assumes a passive relationship between leader and follower, with the latter interested to select the former by homophily. 
The issue of whether leaders might have an active role in changing the preferences of their followers by interacting with them is not discussed at all in this paper; 
in other words, we consider the time scale of eventual changes of preferences to be much larger than the lifespan of a leader-follower interaction, so that users' preferences remain unaltered. 
Extending our model to include more interactive relationships constitutes a fascinating research direction that we will pursue in the future.

\section*{Acknowledgments}
This work was partially supported by the Future and Emerging Technologies programme of the European Commission FP7-COSI-ICT (project QLectives, grant no. 231200), 
by the Swiss National Science Foundation (grant no. 200020-121848), by the National Natural Science Foundation of China (grant nos. 60973069, 90924011 and 60903073) 
and by the International Scientific Cooperation and Communication Project of Sichuan Province in China (grant no. 2010HH0002). 

%

\end{document}